\documentclass[final,5p,times,twocolumn]{elsarticle}

\usepackage{amssymb}
\usepackage[colorlinks]{hyperref}
\hypersetup{hidelinks}

\journal{Nuclear Inst. and Methods in Physics Research, A}

\begin{document}

\begin{frontmatter}



\title{Particle identification with the cluster counting technique for the IDEA drift chamber}


\author[inst1]{C. Caputo}
\author[inst3]{G. Chiarello}
\author[inst3]{A. Corvaglia}
\author[inst3,inst4]{F. Cuna}
\author[inst5,inst6]{B. D'Anzi\corref{cor1}}
\ead{brunella.danzi@ba.infn.it}
\author[inst6,inst7]{N. De Filippis}
\author[inst6]{W. Elmetenawee}
\author[inst3,inst4]{E. Gorini}
\author[inst3]{F. Grancagnolo}
\author[inst3,inst4]{M. Greco}
\author[inst8]{S. Gribanov}
\author[inst9]{K. Johnson}
\author[inst6]{M. Maggi}
\author[inst3]{A. Miccoli}
\author[inst3,inst4]{M. Panareo}
\author[inst8]{A. Popov}
\author[inst3]{M. Primavera}
\author[inst1]{A. Taliercio}
\author[inst6]{G. F. Tassielli}
\author[inst3]{A. Ventura}
\author[inst10,inst11]{S. Xin}

\cortext[cor1]{Corresponding author.}

\affiliation[inst1]{organization={Institut de recherche en mathématique et physique, Université Catholique de Louvain},
            addressline={Chemin du Cyclotron 2/L7.01.01}, 
            city={Louvain-la-Neuve},
            postcode={1348}, 
            country={Belgium}}

\affiliation[inst3]{organization={Istituto Nazionale di Fisica Nucleare},
	addressline={Via Arnesano, 0}, 
	city={Lecce LE},
	postcode={73100}, 
	country={Italy}}

\affiliation[inst4]{organization={Università del Salento},
	addressline={Piazza Tancredi, 7}, 
	city={Lecce LE},
	postcode={73100}, 
	country={Italy}}

\affiliation[inst5]{organization={Università degli Studi di Bari Aldo Moro},
	addressline={Piazza Umberto I, 1}, 
	city={Bari BA},
	postcode={70121}, 
	country={Italy}}

\affiliation[inst6]{organization={Istituto Nazionale di Fisica Nucleare},
	addressline={Via Giovanni Amendola, 173}, 
	city={Bari BA},
	postcode={70126}, 
	country={Italy}}

\affiliation[inst7]{organization={Politecnico di Bari},
	addressline={Via Edoardo Orabona, 4}, 
	city={Bari BA},
	postcode={70126}, 
	country={Italy}}

\affiliation[inst8]{organization={No affiliation},
	country={Russian Federation}}

\affiliation[inst9]{organization={Florida State University},
	addressline={600 W College Ave}, 
	city={Tallahassee FL},
	postcode={32306}, 
	country={United States}}

\affiliation[inst10]{organization={Chinese Academy of Sciences (CAS)},
	addressline={19A Yuquan Road, Shijing District}, 
	city={Beijing},
	postcode={100049}, 
	country={China}}
\affiliation[inst11]{
	organization={Institute of High Energy Physics}, 
	addressline={19B, Yuquan Road, Shijing District},
	city={Beijing},
	postcode={100049}, 
	country={China}}

\begin{abstract}
IDEA (Innovative Detector for an Electron-positron Accelerator) is a general-purpose detector concept, designed to study electron-positron collisions in a wide energy range in a very large circular leptonic collider. Its drift chamber is designed to provide an efficient tracking, a high precision momentum measurement and an excellent particle identification by exploiting the application of the cluster counting technique. To investigate the potential of the cluster counting techniques on physics events, a simulation of the ionization cluster generation is needed, therefore we developed an algorithm which can use the energy deposit information provided by the Geant4 toolkit to reproduce, in a fast and convenient way, the cluster number and cluster size distributions. The results obtained confirm that the cluster counting technique allows to reach a resolution two times better than the traditional dE/dx method. A beam test has been performed during November 2021 at CERN on the H8 beam line to validate the simulations results, to define the limiting effects for a fully efficient cluster counting and to count the number of electron clusters released by an ionizing track at a fixed $\beta\gamma$ as a function of the track angle. The simulation and the beam test results will be described briefly in this issue.
\end{abstract}



\begin{keyword}
particle identification \sep cluster counting \sep drift chamber \sep IDEA detector concept
\end{keyword}
\end{frontmatter}
\section{Introduction}\label{sec:sample1}
The information about ionization energy loss (dE/dx) by a charged particle track in a gaseous detector is traditionally exploited to perform the particle identification (PID) at modern High Energy physics colliders. However, the large and intrinsic uncertainties in the total energy deposition represent a limit to the particle separation capabilities. On the other hand, the digital nature of the cluster counting (CC) technique (dN$_{cl}$/dx) represents a valid alternative to the truncated mean of the energy loss (dE/dx) method, inherently analogical, since it takes advantage of the Poisson nature of the primary ionization process and offers a more statistically robust procedure to infer mass information. This is particularly true for He-based drift chambers with low density of the primary ionization clusters. The technique consists in singling out, in ever recorded detector signal, the isolated structures related to the arrival on the anode wire of the electrons belonging to a single primary ionization act. The goodness of this method relies on the number of primary ionizations being independent from cluster size and gas gain fluctuations and being insensitive to highly ionizing $\delta$-rays. It is going to be used in the high granularity, low mass, fully stereo, co-axial, He-based, unique-volume with the 2 T solenoid field drift chamber tracking system of the IDEA detector concept \cite{Blondeletal2008,Fabioetal2013,FabricioLiang2013}. The choice of a He-based gas mixture implies low drift velocity (of  $\approx$ 2.5 cm/$\mu$s), a larger cluster time separation (30 ns in 90\% He), a low average cluster size $<N_{electrons}/cluster >$  $\approx$ 1.6, and a low single electron diffusion ($< 4.5$ ns for 0.5 cm drift) \cite{Vehlowetal2013,NewmanGirvan2004}.
The effectiveness of the CC algorithms for PID has been demonstrated by theoretical results and presented in \autoref{sec:simulation}. In \autoref{sec:testbeam}, we describe briefly the application of two CC algorithms to real beam test data and their performance results. Taking into account the good outcome, we present in \autoref{sec:conclusion} the future plan for optimizing and validating the CC algorithms and the CC technique for PID.
\section{Analytical calculations and Simulations}\label{sec:simulation}
\begin{figure}
		\centering
		\includegraphics[scale=0.25]{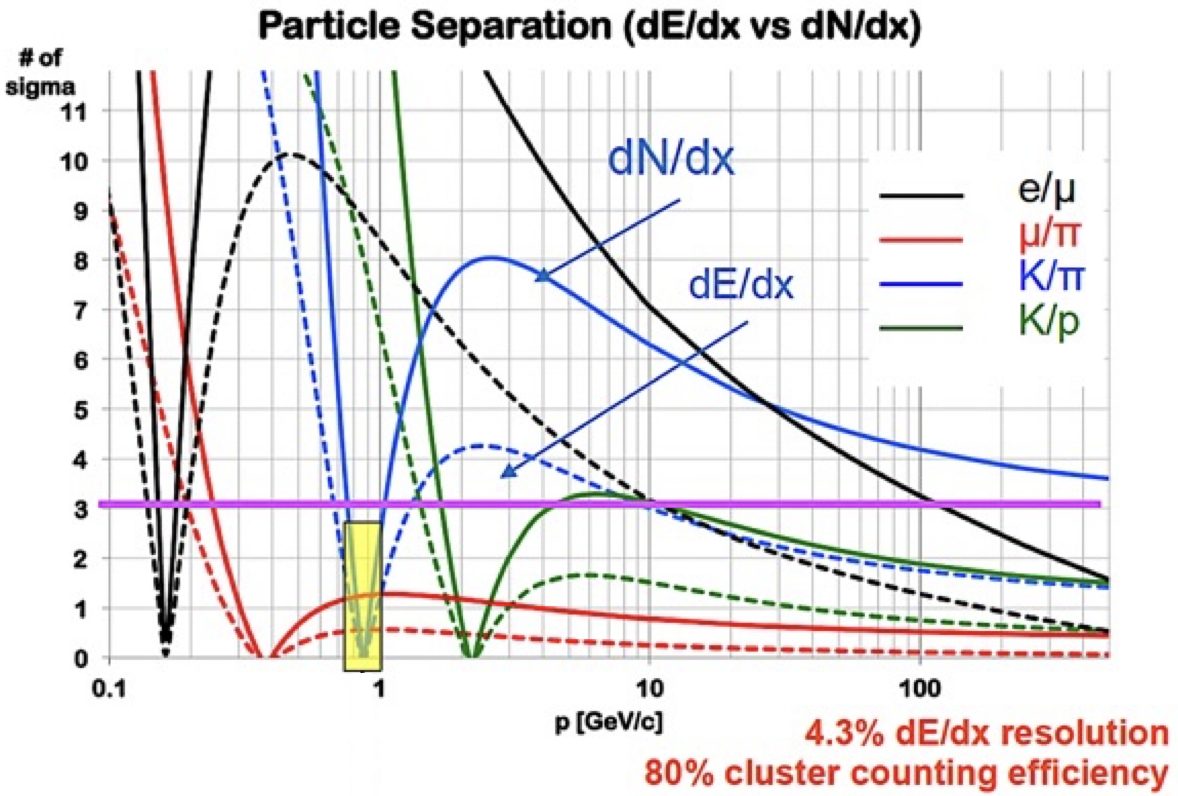}
		\caption{Analytic evaluation of particle separation capabilities achievable with dE/dx (dashed curves) and dN/dx (solid curves). The region between 0.85 GeV/c and 1.05 GeV/c, where a different technique is needed, is highlighted in yellow.} \label{fig:analytic}
	\end{figure}

\begin{figure}
	\centering
	\includegraphics[scale=0.342]{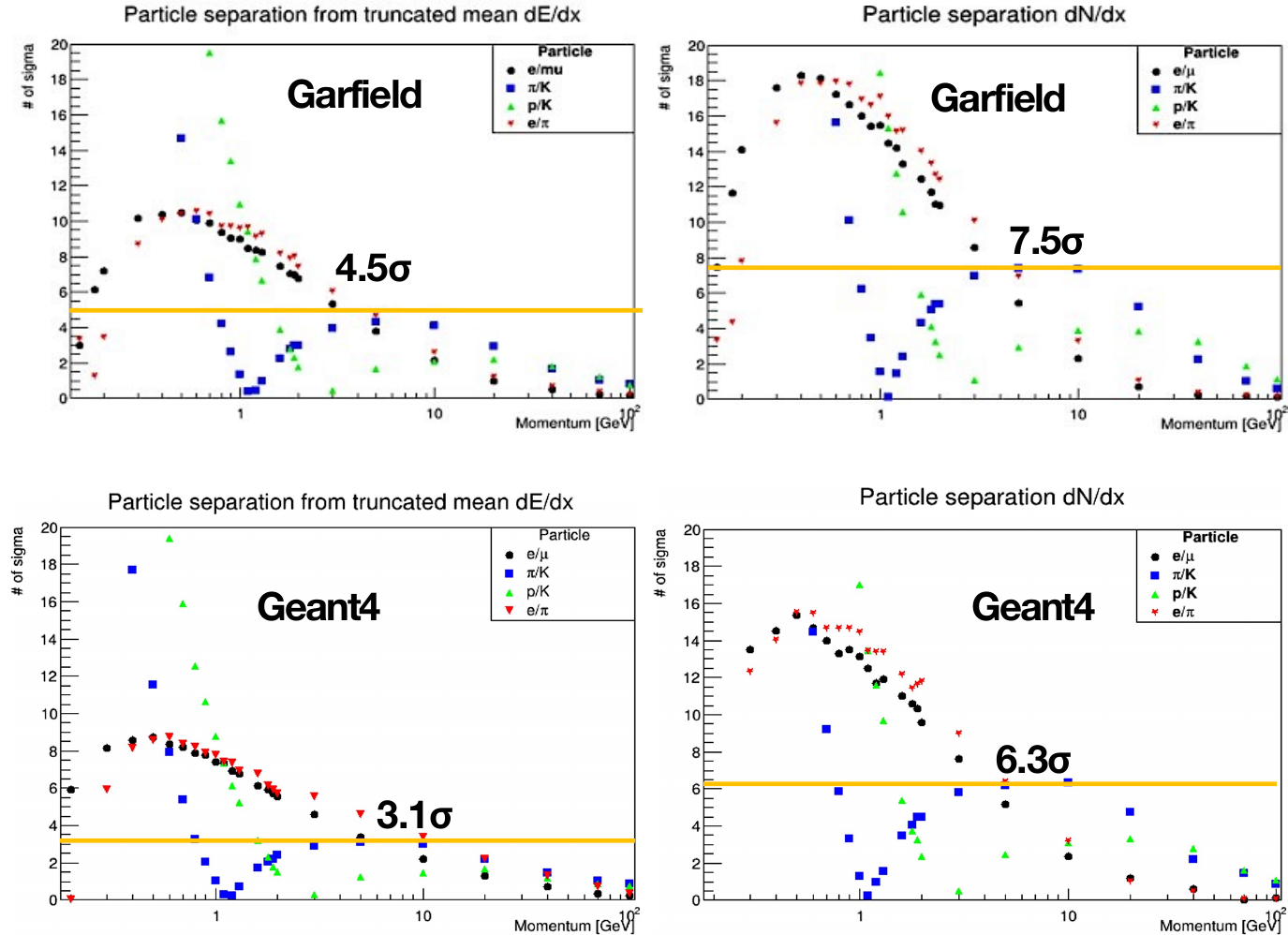}
	\caption{Particle separation capability results from a simulation of the ionization process in 200 drift cells, 1 cm wide, in He/iC$_4$H$_{10}$ 90/10 gas mixture that has been performed both in Garfield++ (top) and in Garfield-modelled Geant4 (bottom) by using the energy loss method (left) and CC technique (right).} \label{fig:simul}
\end{figure}

From analytical calculations, we infer an excellent $K/\pi$ separation for He/iC$_4$H$_{10}$ 90/10 gas mixture over the entire momentum range except for $0.85 < p < 1.05$ GeV (blue lines in \autoref{fig:analytic}), which can be easily recovered with a modest (100 ps resolution over approximately 2 m path length) time of flight measurement. On the other hand, starting from detailed studies of Garfield++ simulation results about the ionization process in He-based gas mixtures, an algorithm which reproduces the cluster number and cluster size distributions using the energy deposit information by Geant4 has been developed under the assumption of a 100\% CC efficiency \cite{Fortunato2010}. 
From \autoref{fig:simul}, we observe that the dN$_{cl}$/dx improves particle separation capabilities and Garfield++ is in good agreement with analytical calculations up to 20 GeV/c, then falls much more rapidly at higher momenta.
\section{Cluster counting technique applied to beam test data}\label{sec:testbeam}
\begin{figure}
	\includegraphics[scale=0.6]{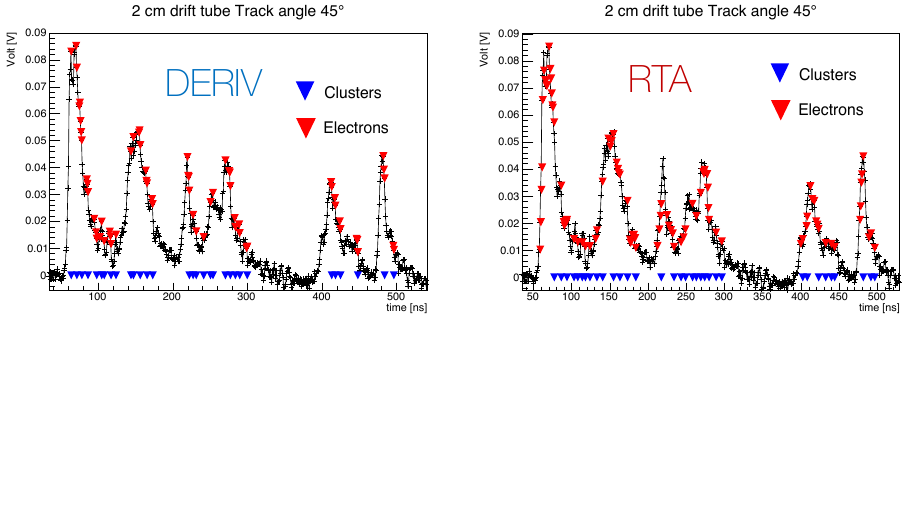}
	\caption{Example of real waveforms collected from 2 cm size drift tubes (90/10 He/iC$_4$H$_{10}$). The blue and red arrows represent the cluster and electron peaks found with the DERIV and RTA algorithms.}  \label{fig:cluster}
\end{figure}

\begin{figure}
	\centering
	\includegraphics[scale=0.26]{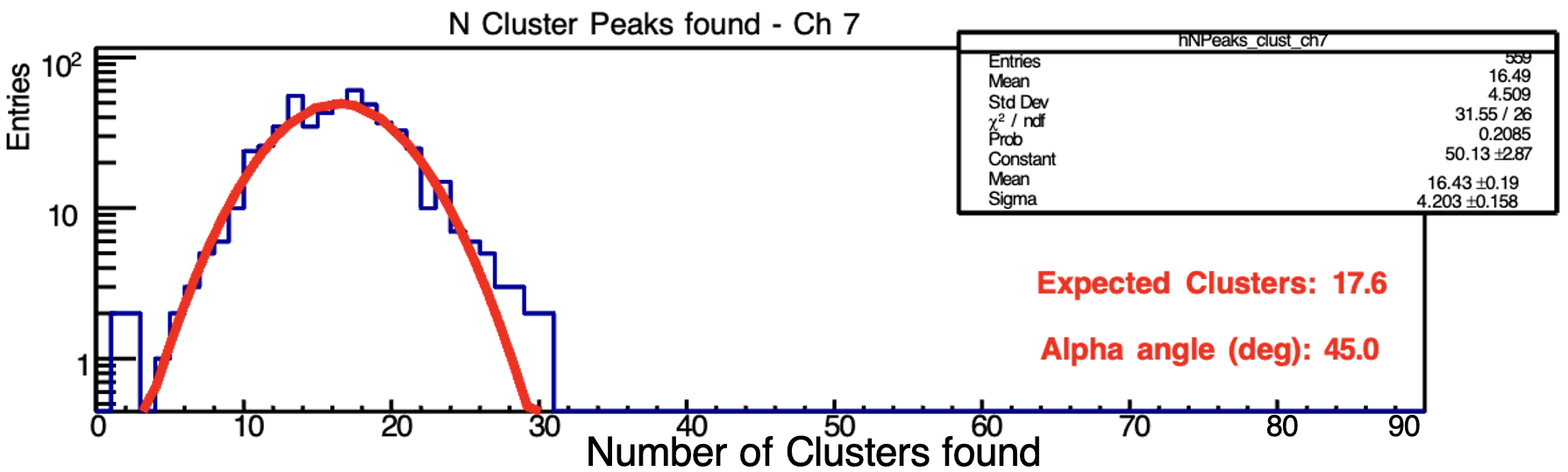}
	\caption{Distribution of the number of clusters for 1 cm size drift tubes in He:IsoB 90/10 fitted with a gaussian.} \label{fig:efficiency}
\end{figure}

To validate the simulations results, a first beam test, using a 165 GeV/c muon beam on a setup made of 1 cm and 2 cm size drift tubes, equipped with Mo, Al, and W sense wires in the 10 - 40 $\mu m$ diameter range, has been performed at CERN/H8 beam line by collecting data with the gas mixtures He/iC$_4$H$_{10}$ 90/10 and 80/20, by making a scan in high voltage and angle between the wire direction and ionizing tracks. Fulfilling the main requirements for an efficient CC on the readout electronics (1 GHz bandwidth and at least a 1 GS/s sampling rate at 12 bits), entirely based on a DRS system \cite{DRS}, has been crucial. The most efficient CC algorithms are the Running Template Algorithm (RTA), scanning the waveform with an electron peak template and its iterative subtraction, and the Derivative Algorithm (DERIV), imposing cuts on the waveform amplitude, the first and second derivatives (see \autoref{fig:cluster}). They define the limiting effects for a fully efficient CC (space charge effect + attachment + recombination) which is approximately 80\% in He/iC$_4$H$_{10}$ 90/10 gas mixture. \autoref{fig:efficiency} shows the distribution of the number of clusters for a particular configuration. The gaussian fit indicates a mean value $\mu$ very close to the expected one and a sigma almost equal to the square root of $\mu$, confirming the Poisson nature of the distribution.
\section{Conclusion and Future Perspectives}
\label{sec:conclusion}
Using preliminary CC algorithms gave promising results in reproducing the expected physical distributions. As a next step, the experimental setup will undergo new tests in muon beams with momenta over the full range of interest for the future lepton machines, to assess the particle identification capabilities of the cluster counting approach with optimized algorithms.

 \bibliographystyle{elsarticle-num} 
 \bibliography{cas-refs}





\end{document}